\documentclass[utf8]{FrontiersinHarvard} 

\usepackage{url,lineno,microtype,subcaption}
\usepackage[onehalfspacing]{setspace}
\usepackage{amsfonts,amsmath,graphicx,braket,bm}
\usepackage[colorlinks, linkcolor=blue, citecolor=blue]{hyperref}

\DeclareMathOperator{\sgn}{sgn}

\def\keyFont{\fontsize{8}{11}\helveticabold }
\def\firstAuthorLast{Díez-Valle {et~al.}} 
\def\Authors{Pablo Díez-Valle\,$^{1,*}$, Diego Porras\,$^{1}$ and Juan José García-Ripoll \,$^{1}$}
\def\Address{$^{1}$Instituto de F\'{\i}sica Fundamental IFF-CSIC, Calle Serrano 113b, Madrid 28006, Spain} 
\def\corrAuthor{Pablo Díez-Valle}

\def\corrEmail{pablo.diez@csic.es}

\begin{document}
\onecolumn
\firstpage{1}

\title {Connection between single-layer Quantum Approximate Optimization Algorithm interferometry and thermal distributions sampling} 

\author[\firstAuthorLast ]{\Authors} 
\address{} 
\correspondance{} 

\extraAuth{}

\maketitle

\begin{abstract}


\section{}

The Quantum Approximate Optimization Algorithm (QAOA) is an algorithm originally proposed to find approximate solutions to Combinatorial Optimization problems on quantum computers. However, the algorithm has also attracted interest for sampling purposes since 
it was theoretically demonstrated under reasonable complexity assumptions that one layer of the algorithm already engineers a probability distribution beyond what can be simulated by classical computers. In this regard, a recent study has shown as well that, in universal Ising models, this global probability distribution resembles pure but thermal-like distributions at a temperature that depends on internal correlations of the spin model. In this work, through an interferometric interpretation of the algorithm, we extend the theoretical derivation of the amplitudes of the eigenstates, and the Boltzmann distributions generated by single-layer QAOA. We also review the implications that this behavior has from both a practical and fundamental perspective.

\tiny
 \keyFont{ \section{Keywords:} quantum computing, quantum algorithms, quantum optimization, adiabatic computing, variational algorithms, statistical physics, machine learning, quantum sampling} 
\end{abstract}

\newcommand{\jjgr}[1]{\textcolor{black}{#1}}

\section{Introduction}
Variational quantum algorithms (VQAs) are a promising framework for solving computationally hard tasks~\citep{Cerezo_2021,Bharti_2022}. \jjgr{These algorithms are based on a quantum circuit with tunable parameters acting as ansatz.} Finding the minimum energy state among a discrete set of possible solutions (combinatorial optimization problems)~\citep{Nemhauser1988IntegerAC,Moll_2018}, or drawing samples from a classical probability distribution (sampling problems)~\citep{Lund_2017,Wild_2021} are two examples of worthwhile challenges \jjgr{addressed in this framework}. 

Among the plethora of proposed VQAs, the Quantum Approximate Optimization Algorithm (QAOA)~\citep{Farhi_2014} has received special attention from the scientific community, with remarkable empirical and theoretical results on the \jjgr{algorithm's performance}~\citep{Harrigan_2021,Farhi_2022,Blekos_2023}. The ansatz of QAOA is inspired by a trotterized adiabatic evolution capable of approximating the minimum energy state of a given cost function. An advantage of the practical implementation of QAOA is that the number of variational parameters in the quantum circuit to be optimized is independent of the problem size, on the other hand, the depth of the circuit increases with the quality of the approximation of the fundamental state.  

\jjgr{Beyond QAOA's original use in approximating ground states, QAOA has been recently shown to have interest from a global sampling perspective}~\citep{Farhi_2019}. It has been shown that the shallowest version of the algorithm produces a probability distribution that cannot be simulated by classical computation since, otherwise, the Polynomial Hierarchy would collapse. Therefore, sampling the QAOA circuit in some variational parameters range could exhibit \textit{quantum supremacy}~\citep{Preskill_2018}, \jjgr{understood as a task that can be more efficiently done in a quantum computer than in a classical one}. This result is connected to previous demonstrations of the hardness of simulating quantum circuits, for instance Boson Sampling~\citep{Brod_2015} or Instantaneous Quantum Polynomial time (IQP) circuits~\citep{Bremner_2010,Bremner_2016}. 

Despite the importance of this theoretical result, it gives no clue about the nature of the probability distribution \jjgr{generated by QAOA, nor whether sampling from it has any} practical interest beyond combinatorial optimization or a quantum supremacy demonstration. Following this direction, in~\citep{Diez_Valle_2023}, we studied the global structure and probability distribution in the energy space of the single-layer QAOA ansatz. In that letter, we provide evidence for a clear connection between the QAOA and thermal distribution sampling with probability amplitude distributions resembling Boltzmann distributions at low temperatures. We call such distributions \textit{QAOA pseudo-Boltzmann states}. The excellent performance of QAOA revealed at temperatures below the state-of-the-art theoretical limit to fast mixing of Markov Chain Monte Carlo methods~\citep{eldan_2021}.

In the current manuscript, we expand the theoretical derivation of pure thermal-like QAOA states introduced in~\citep{Diez_Valle_2023}. The paper is structured as follows. First, in Section~\ref{sec:Interferometric interpretation of QAOA}, we review the main ingredients of the Quantum Approximate Optimization Algorithm and introduce an intuitive picture of QAOA as an interferometer in energy space. In Section~\ref{subsec:General scenario} we extend the interferometric picture to the multilayer scenario and derive the analytical amplitudes of the final wavefunction. In Section~\ref{subsec:Ising Models}, we specify the results of the previous section to the case where an Ising Model Hamiltonian defines the cost function. We detail the internal correlations between the eigenstates of the Hamiltonian in degenerate and nondegenerate scenarios, leading to a closed expression for the probability amplitudes of the QAOA wavefunction. We connect such probability amplitudes to the sampling of Boltzmann or thermal-like distributions in Section~\ref{sec:QAOA thermal-like distributions}, also analyzing the features that these results reveal about the distribution behavior as we change the variational parameters of the algorithm. We conclude with a perspective on the importance of the pseudo-Boltzmann QAOA states and approximate thermal sampling in Section~\ref{sec:Outlook}.

\section{Interferometric interpretation of QAOA}
\label{sec:Interferometric interpretation of QAOA}

\subsection{The Quantum Approximate Optimization Algorithm}
The Quantum Approximate Optimization Algorithm (QAOA) \jjgr{aims to} find good approximate solutions to combinatorial optimization problems \citep{Farhi_2014}, \jjgr{which are} defined by a classical objective function $E(\mathbf{x})$ mapping $N$-binary strings to real values:
\begin{equation}
 E(\mathbf{x}) :\{0,1\}^N \longrightarrow \mathbb{R}
\end{equation}
The algorithm's goal is to find a binary string $\mathbf{x}^*$ that minimizes the function $E_{min} \equiv \min_{\mathbf{x}}E(\mathbf{x})$, or at least that achieves a good approximation ratio $r\in[0,1]$
\begin{equation}
    \dfrac{\max_{\mathbf{x}}E(\mathbf{x}) - E(\mathbf{x})}{\max_{\mathbf{x}}E(\mathbf{x})-\min_{\mathbf{x}}E(\mathbf{x})} \geq r \,.
\end{equation}
\jjgr{This optimization problem is equivalent to finding the ground state of a spin Hamiltonian $\hat{E}$, where each binary variable $x_i$ describes the state of one spin $\sigma^z_i$ as $\mathbf{x} \rightarrow \frac{1}{2}(1+\bm{\sigma}^z)$,}
\begin{equation}
\hat{E}_{min} \equiv \min_{\bm{\sigma}^z} \bra{\bm{\sigma}^z} \hat{E}(\bm{\sigma}^z) \ket{\bm{\sigma}^z} =  \min_{\mathbf{x}} \bra{\mathbf{x}} \hat{E}(\mathbf{x}) \ket{\mathbf{x}}\,.
\end{equation}
 
A well-known approach to solving these problems is the Quantum Adiabatic Algorithm (QAA), which ensures the achievement of the global minimum given a sufficiently long run time $T$ \citep{farhi2000quantum, Farhi_2001}. The adiabatic protocol is guided by \jjgr{a time-dependent Hamiltonian such as}
\begin{equation}
    \hat{H}_{\textnormal{QAA}}(t) = -\left[\left(1-f\left(\frac{t}{T}\right)\right) \hat{H}_x + f\left(\frac{t}{T}\right) \hat{E}\right]\, 
\end{equation}
where $\hat{H}_x = \sum_i^N \sigma^x_i$ with $\sigma^x_i$ the Pauli $X$ operator acting on the ith qubit, and $f(x)$ is the schedule function with $f(0)=0$ and $f(1)=1$. The adiabatic theorem guarantees that a sufficiently slow time evolution will keep the system in its ground state. \jjgr{Thus, an evolution that starts with the ground state of} $\hat{H}_x$, 
\begin{equation}
\ket{\Psi(0)} = \frac{1}{\sqrt{2^N}} \sum_{\mathbf{x}}\ket{x} \equiv \ket{+}^{\otimes N},  
\end{equation} 
will approximately bring the system to the ground state of $\hat{E}$, 
\begin{equation}
    \ket{\Psi(T)} = \ket{\mathbf{x}_{min}} .
\end{equation}
To ensure success, the runtime $T$ must scale as $T = O\left(\Delta^{-2}_{\textnormal{min}}\right)$, where $\Delta_{\textnormal{min}}$ is the minimum spectral gap~\citep{Albash_2018}. 
QAOA proposes a trotterized approximation to adiabatic evolution consisting of a quantum circuit built by the alternation of the following two operators
\begin{equation}
    \ket{\Psi_{\textnormal{QAOA-p}}(\bm{\gamma},\bm{\theta})} = U(\hat{H}_x,\theta_p) U(\hat{E},\gamma_p) ... U(\hat{H}_x,\theta_2) U(\hat{E},\gamma_2)U(\hat{H}_x,\theta_1) U(\hat{E},\gamma_1) H^{\otimes N}\ket{0}^{\otimes N}\,, 
\label{eq_qaoaansatz}
\end{equation}
with $H$ denoting single-qubit Hadamard gates, and 
\begin{equation}
    U(\hat{E},\gamma) = e^{-i\gamma\hat{E}}   
\end{equation}
\begin{equation}
    U(\hat{H}_x,\theta) = e^{-i\theta\hat{H}_x} = \prod_j^N e^{-i\theta\sigma^x_j} \equiv R_x(\theta)^{\otimes N}   
\end{equation}
where $\bm{\gamma} = (\gamma_1,\gamma_2,...,\gamma_p)$ and $\bm{\theta} = (\theta_1,\theta_2,...,\theta_p)$ are set of variational angles that can be tuned to approximate the ground state of the cost Hamiltonian $\hat{E}$, and $p$ is the number of layers that determines the depth of the quantum circuit and the accuracy of the algorithm. 
The infinite depth limit $p\rightarrow\infty$ returns the adiabatic evolution showing that we can get a good enough approximation to the minimum of the optimization problem for sufficiently large $p$.   

\subsection{QAOA as an interferometer in energy space}
\jjgr{Let us highlight QAOA's potential} in small-depth regimes by introducing an \jjgr{interpretation QAOA's circuit as an interferometer} operating globally in energy space. \jjgr{We will derive analytical results for the interference amplitude in later sections, focusing} on the $p=1$ single-layer ansatz (see Eq.~\eqref{eq_qaoaansatz}),
\begin{figure}
\begin{center}
\includegraphics[width=0.7\textwidth]{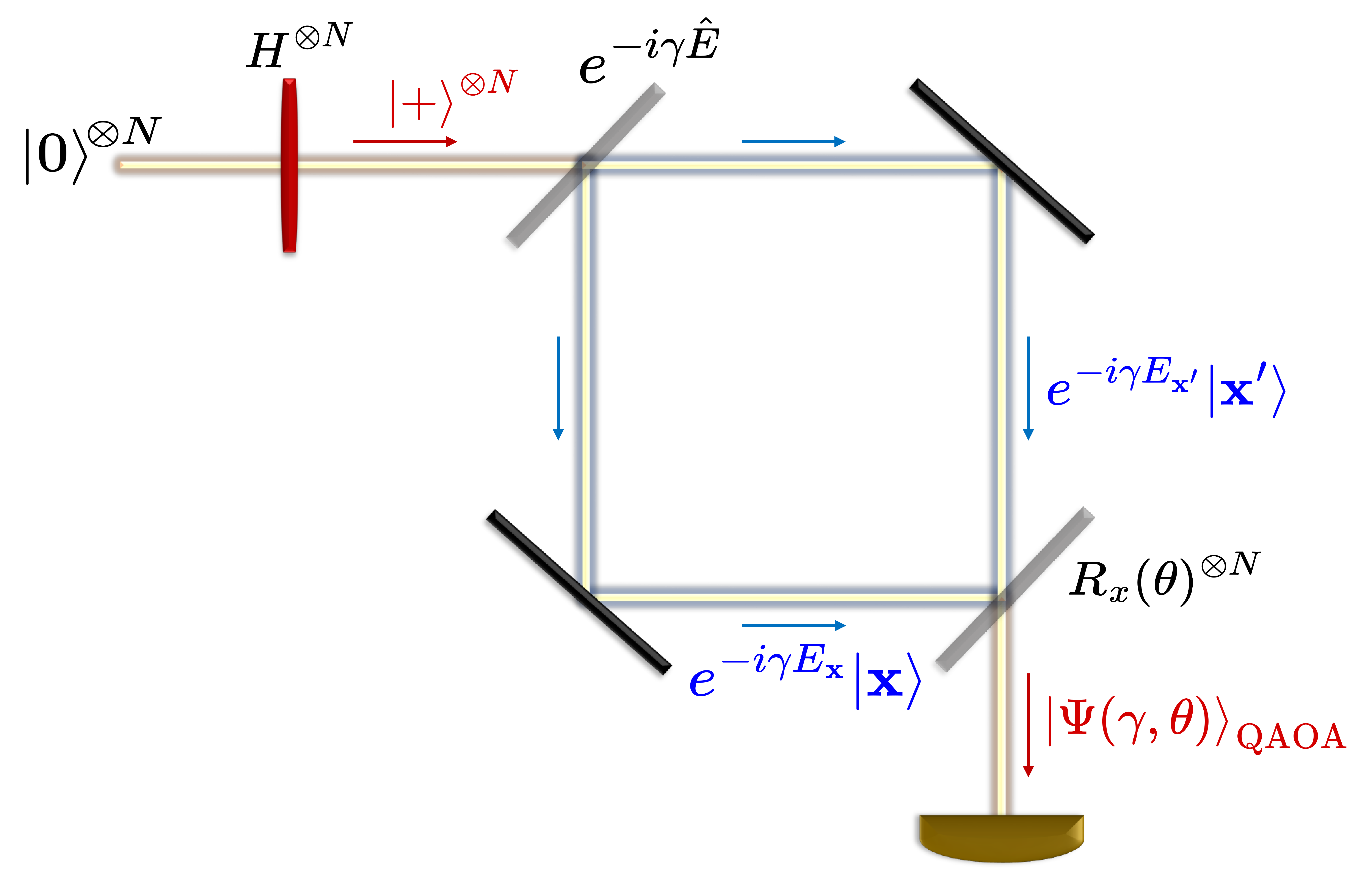}
\end{center}
\caption{Interferometric interpretation of the Quantum Approximate Optimization Algorithm circuit.}\label{fig:qaoa_interferometer}
\end{figure}
\begin{equation}
    \ket{\Psi_{\textnormal{QAOA-1}}(\gamma,\theta)} = R_x(\theta)^{\otimes N} e^{-i\gamma \hat{E}} H^{\otimes N}\ket{0}^{\otimes N}.
\end{equation}

As sketched in Fig.\ref{fig:qaoa_interferometer}, \jjgr{the Hadamard gate splits the quantum state $\ket{\mathbf{0}}$ into a superposition of all states in the computational basis, acting like a multidimensional "mirror"}
\begin{equation}
     |s_1\rangle = \ket{+}^{\otimes N} = \dfrac{1}{\sqrt{2^N}} \sum_{\mathbf{x}}\ket{\mathbf{x}}\,.
\end{equation}
\jjgr{Then the evolution with the diagonal cost Hamiltonian $\hat{E}\ket{\mathbf{x}} = E_\mathbf{x} \ket{\mathbf{x}},$ given by the operator $U_E(\hat{E},\gamma)$, imparts phases on all states, acting like the branches of the interferometer}
\begin{equation}
    |s_2\rangle = e^{-i\gamma \hat{E}} |s_1\rangle =  \dfrac{1}{\sqrt{2^N}}  \sum_{\mathbf{x}} e^{-i\gamma E_\mathbf{x}} \ket{\mathbf{x}} .  
\end{equation}
Finally, the mixing operator $R_x(\theta)^{\otimes N}$ recombines the energy states so that the interference transforms the energy-dependent relative phases in measurable probability amplitudes
\begin{equation}
    \ket{\Psi_{\textnormal{QAOA-1}}(\gamma,\theta)} = R_x(\theta)^{\otimes N} |s_2\rangle  = \sum_\mathbf{x} F_{\mathbf{x}}(E_{\mathbf{x}},\gamma,\theta) \ket{\mathbf{x}} .
\end{equation}

\jjgr{This intuition is much clearer when looking at the algorithm operating on a single qubit, with rotation} $R_x(\theta)$,
\begin{equation}
    R_x(\theta) = e^{-i\theta\sigma^x} = \mathbb{I}\cos\theta - i\sigma^x\sin\theta  = \begin{pmatrix} \cos\theta & -i\sin\theta\\
    -i\sin\theta & \cos\theta \end{pmatrix} \,. 
    \label{rx}
\end{equation}
Consider the following two-state subspace  
$$\left\{|\mathbf{x}_0\rangle = |0,x_1,x_2,...,x_N\rangle ,|\mathbf{x}_1\rangle = |1,x_1,x_2,...,x_N\rangle\right\},$$ 
where $x_i=\{0,1\}$ (so $\mathbf{x}_0$ and $\mathbf{x}_1$ differ by only one bit), with energies $E_{\mathbf{x}_0} \equiv \bra{\mathbf{x}_0} \hat{E} \ket{\mathbf{x}_0}$, $E_{\mathbf{x}_1} \equiv \bra{\mathbf{x}_1} \hat{E} \ket{\mathbf{x}_1}$. The probability amplitudes before the local interference are
\begin{equation}
    A_{\mathbf{x}_0}^2 = |\braket{ \mathbf{x}_0|s_2}|^2\,,\,  A_{\mathbf{x}_1}^2 = |\braket{ \mathbf{x}_1|s_2}|^2 ,
\end{equation} 
respectively. After applying the local gate in Eq.~\eqref{rx} on the qubit which corresponds to the different bit, the interference shifts the amplitudes to $N_{\mathbf{x}_0}$ and $N_{\mathbf{x}_1}$,
\begin{equation}
    \begin{pmatrix} N_{\mathbf{x}_0} \\
        N_{\mathbf{x}_1}
    \end{pmatrix} = R_x(\theta) \begin{pmatrix} e^{-i\gamma E_{\mathbf{x}_0}}A_{\mathbf{x}_0}  \\
       e^{-i\gamma E_{\mathbf{x}_1}} A_{\mathbf{x}_1} 
    \end{pmatrix} = e^{-i\gamma E_{\mathbf{x}_0}} \begin{pmatrix} A_{\mathbf{x}_0}\cos\theta-i A_{\mathbf{x}_1}e^{-i\gamma F_{\mathbf{x}_1,\mathbf{x}_0}}\sin\theta  \\
    -i A_{\mathbf{x}_0}\sin\theta+A_{\mathbf{x}_1}e^{-i\gamma F_{\mathbf{x}_1,\mathbf{x}_0}}\cos\theta 
    \end{pmatrix} \,,
\end{equation}
where $F_{\mathbf{x}_1,\mathbf{x}_0}\equiv E_{\mathbf{x}_1} - E_{\mathbf{x}_0}$. Thus, the probability amplitudes become
\begin{equation}
\begin{split}
N_{0}^2 = & A_{\mathbf{x}_0}^2 \cos^2\theta + A_{\mathbf{x}_1}^2 \sin^2\theta - A_{\mathbf{x}_0}A_{\mathbf{x}_1} \sin\left(2\theta\right) \sin\left(\gamma F_{\mathbf{x}_1,\mathbf{x}_0}\right) \\ 
N_{1}^2 = & A_{\mathbf{x}_0}^2 \sin^2\theta + A_{\mathbf{x}_1}^2 \cos^2\theta + A_{\mathbf{x}_0}A_{\mathbf{x}_1} \sin\left(2\theta\right) \sin\left(\gamma F_{\mathbf{x}_1,\mathbf{x}_0}\right)  \,.
\label{eq_newamplitudes}
\end{split}  
\end{equation}
Note that the relative phase between the states $\gamma F_{\mathbf{x}_1,\mathbf{x}_0}= \gamma \left(E_{\mathbf{x}_1} - E_{\mathbf{x}_0}\right)$ controls the amplification. Therefore, assuming $\sin\left(2\theta\right)<0$ and $\gamma>0$ or vice versa and provided the angle $\gamma$ is sufficiently small, the interference term always increases the population of the lowest energy state and reduces that of the other. Furthermore, Eq. \eqref{eq_newamplitudes} also reflects how the relative sign between $\sin\left(2\theta\right)$ and $\gamma$ controls the direction of optimization. The probability amplitude of the lowest/highest energy state is enhanced when the signs are opposite/equal.

\begin{figure}[]
\begin{center}
\includegraphics[width=0.8\textwidth]{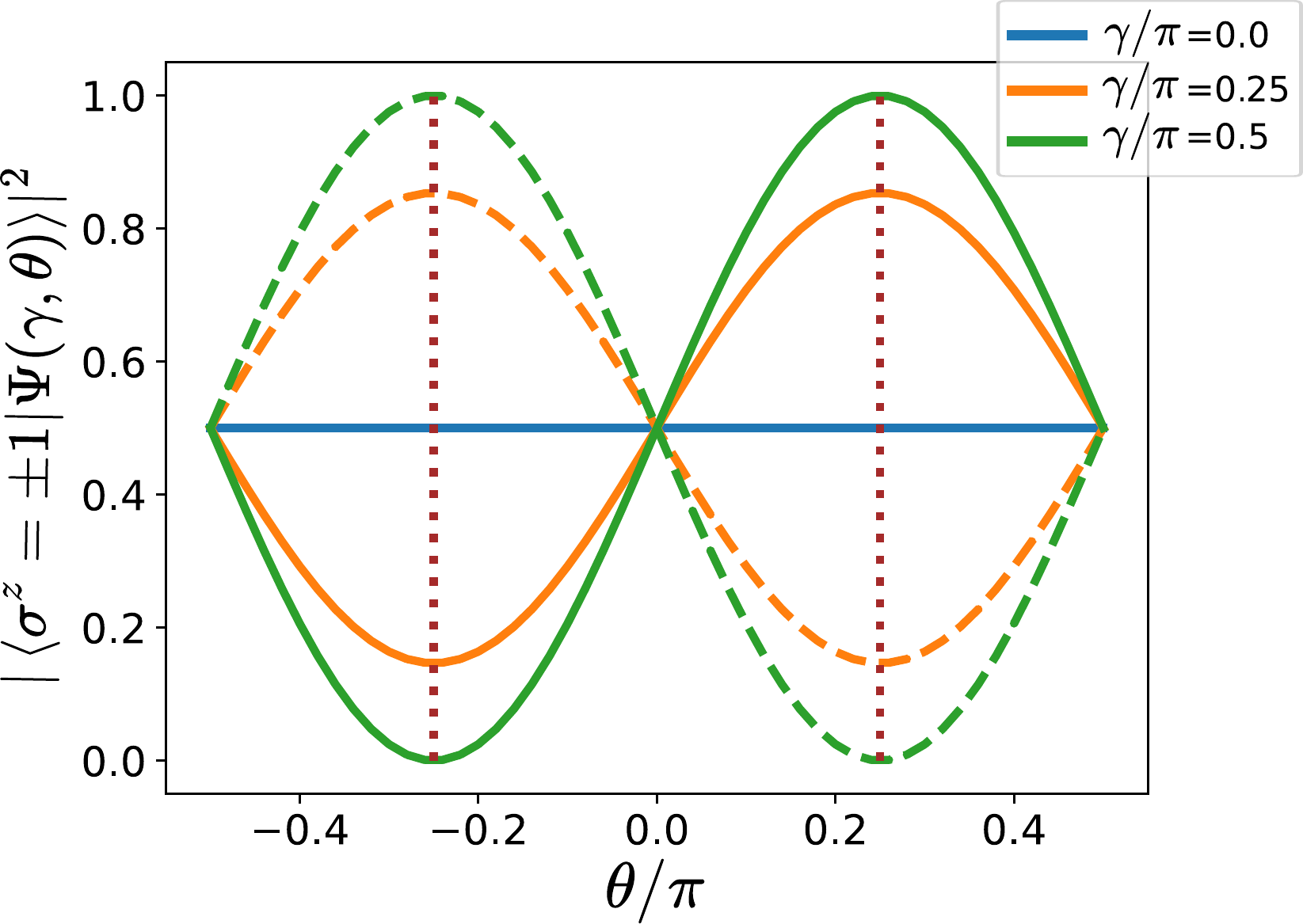}
\end{center}
\caption{Quantum Approximate Optimization Algorithm interferometry on the two-level spin system defined by the Hamitonian $\hat{E}=\frac{1}{2}\Delta\sigma^z$ with $\Delta=1$. We plot the probability amplitude of the ground state $\sigma^z = -1$ (dashed lines), and the highest energy state $\sigma^z = +1$ (solid lines) for different values of the QAOA angles $\theta$ and $\gamma$. We highlight the points of maximum amplification $\theta=\pm\pi/4$, $\gamma=\pi/2$ (dotted brown lines).}\label{fig:twolevel_interferometry}
\end{figure}

This interferometric behavior expanded to the $N$-level scenario imposes nontrivial, structure-dependent upper bounds on the $|\theta|$ and $|\gamma|$ angles to avoid symmetries and random scrambling of the energy states. In particular, Eq.~\eqref{eq_newamplitudes} shows that for the single-qubit interference 
\begin{equation}
  n\pi\leq\theta<(n+1)\pi \,\,\,\textnormal{and}\,\,\, |\gamma|<\frac{\pi}{F_{\mathbf{x}_{max},\mathbf{x}_{min}}} \sim O(||\hat{E}||^{-1})\,,   
\end{equation}
with \jjgr{$n\in\mathbb{Z}$}. The $N$-qubit interference makes this picture richer and more complicated as derived in the next sections. Analytical and numerical results for the single layer QAOA on general Ising spin models show that the optimal $|\gamma|$ for an $N$ qubit system actually scales as $O(N \left(N|e|\right)^{-1/2})$~\citep{Ozaeta_2022,Diez_Valle_2023}, where $|e|$ is the number of edges or non-null elements in the coupling matrix $\mathbf{J}$ (see Eq.~\eqref{eq_isinghamiltonian}). For instance, in the two-level system $\hat{E}=\frac{1}{2}\Delta\sigma^z$ the probability of measuring the ground state $\ket{\sigma^z=-1}$ and the highest energy state $\ket{\sigma^z=+1}$ is 
\begin{equation}
    |\braket{ \sigma^z=\pm 1 |\Psi(\gamma,\theta)}|^2 = \frac{1}{2}\left(1 + \sigma^z \, \sin(2\theta)\sin\left(\gamma \Delta\right)\right) \,,
\end{equation}
which is maximal with optimal angles $\theta=\pm\frac{\pi}{4}$ and $\gamma=\frac{\pi}{2\Delta}$ (see Fig.\ref{fig:twolevel_interferometry}). 

\section{Multi-level interferometer amplitude}
\label{sec:Multi-level interferometer amplitude}
\subsection{General scenario}
\label{subsec:General scenario}
\jjgr{Let us extend the two-level picture, to a general framework dealing with} the complete interference spectrum of a single-layer QAOA circuit. We focus on how the Quantum Approximate Optimization Algorithm transforms the wavefunction amplitudes
\begin{equation}
    F(\mathbf{x}) \equiv \braket{\mathbf{x}| \Psi_{\textnormal{QAOA-1}}(\gamma,\theta)}.
    \label{eq_multilevelinterfampl}
\end{equation}
\jjgr{As previously shown, a local $R_x$ rotation acting on the i-th qubit mixes the amplitude between each state $\mathbf{x}$ with the corresponding state $\mathbf{x}'$ that differs only on the value of the i-th bit (see Eq.~\eqref{eq_newamplitudes}). Thus, the action of the whole mixing operator $R_{x}^{\otimes N}$ is that of a complete mixing between all states in the computational basis}
\begin{equation}
    F(\mathbf{x}) =  \sum_{\mathbf{x'}} \frac{e^{-i\gamma E_{\mathbf{x'}}}}{2^{N/2}} \bra{\mathbf{x}} \prod_j^N \left(\cos\theta - i\sigma^x_j \sin\theta\right) \ket{\mathbf{x'}}.
\end{equation}
\jjgr{The weight of each mixing process depends on the number of bits that must be flipped to go from one state $\mathbf{x}$ to the other $\mathbf{x}'$---i.e., the Hamming distance between both states---, and also on the oscillating terms induced by the angles $\theta$ and $\gamma$}. The amplitude of the state generated by the algorithm after the mixing operator $U(\hat{H}_x,\theta)$ is given by an interference formula, 
\begin{equation}
F(\mathbf{x}) \equiv \braket{\mathbf{x}|\tilde\Psi} = \frac{1}{2^{N/2}}\sum_{\mathbf{x'}} \cos\left(\theta\right)^{N-H_{\mathbf{x},\mathbf{x'}}}\left[-i \sin\left(\theta\right)\right]^{H_{\mathbf{x},\mathbf{x'}}}
e^{-i\gamma E_{\mathbf{x'}}}\,
\label{eq_interferenceamplitudes}
\end{equation}
where $H_{\mathbf{x},\mathbf{x'}}$ is the Hamming distance between two bit configurations $\mathbf{x}$ and $\mathbf{x'}$ that represent eigenstates of a spin model with energy $E_\mathbf{x}$ and $E_\mathbf{x'}$. We can unify the weight sum terms into a single exponential by the following change of variables in which the rotation angle $\theta$ is reparameterized in terms of an exponent $r$ and a normalization $R$,
\begin{equation}
\cos\left(\theta\right) = R^{\frac{1}{2}} \exp\left[\frac{r}{2}\right], \; \sin\left(\theta\right) = R^{\frac{1}{2}}\exp\left[-\frac{r}{2}\right],
\label{eq_cosin}
\end{equation}
where due to the symmetries of the interference we consider $\theta\in(0,\frac{\pi}{2})$ without loss of generality. The sign of $\gamma$ allows us to define if the interference increases the population of lower or higher energy states so that we cover all the possibilities. By this transformation, the interference amplitudes in Eq.~\eqref{eq_interferenceamplitudes} become a sum of exponentials over the entire configuration space,    
\begin{equation}
F(\mathbf{x}) = \left(\frac{R\exp [r]}{2}\right)^{\frac{N}{2}}\sum_{\mathbf{x'}}\exp\left[-H_{\mathbf{x}\mathbf{x'}}\left(i\frac{\pi}{2} + r\right) - i\gamma E_{\mathbf{x'}}\right].
\label{eq_qaoaamplexp}
\end{equation}

Since Eq.~\eqref{eq_qaoaamplexp} encompasses all eigenstates of the system encoded in its energy $E_\mathbf{x'}$ and Hamming distance $H_{\mathbf{x},\mathbf{x'}}$ with the reference state $\mathbf{x}$, it is convenient to introduce a probability distribution relating distances in the computational base to the energy spectrum, 
\begin{equation}
  p(H, E; \mathbf{x})
   = \frac{1}{2^{N}}\sum_{\mathbf{x'}} \delta(H - H_\mathbf{xx'}) \delta(E - E_\mathbf{x'}),
\label{eq_probHE}
\end{equation}
This expression represents the relative number of eigenstates that, given a reference state $\mathbf{x}$, have a Hamming distance $H$ to that state and energy equal to $E$. In other words, this distribution captures the internal correlations inherent to the specific spin model. With this definition at hand, we can express the previous sum in Eq.~\eqref{eq_qaoaamplexp} as the average over this probability distribution
\begin{equation}
  F(\mathbf{x}) \propto \left(\exp [r]\right)^{\frac{N}{2}}\int\int_{-\infty}^{\infty}
  \exp\left[-H\left(i\frac{\pi}{2} + r\right) - i\gamma E\right] p(H,E; \mathbf{x}) dHdE.
  \label{eq_amplint}
\end{equation}
Hence, the interference amplitude in Eq.~\eqref{eq_multilevelinterfampl} depends on the structure of the energy levels of the spin system manifested in the probability distribution $p(H, E;\mathbf{x})$. In the following, we study such a structure to derive a unified expression of the QAOA probability amplitudes for universal Ising spin models.   

\subsection{Ising Models}
\label{subsec:Ising Models}

Next, we focus on a universal set of nondeterministic polynomial time hard (NP-Hard) Ising Models~\citep{Barahona_1982} and derive this scenario's interference probability amplitude formula. 

A broad spectrum of combinatorial optimization problems can be represented as a spin model described by an Ising Hamiltonian~\citep{Lucas_2014},
\begin{equation}
    E_I(\mathbf{s}) = \frac{1}{2} \sum_{i,j=1}^N s_i  J_{ij} s_j + \sum_{i=1}^N h_i s_i,
  \label{eq_isinghamiltonian}
\end{equation}
where $\mathbf{s} = \{-1,+1\}^N$, $N$ is the number of variables, $\mathbf{J}$ is an $N$-by-$N$ square coupling matrix, and the magnetic field $\mathbf{h}$ is a vector of $N$ coefficients. The quantum version of this Hamiltonian $\hat{E}_I(\bm{\sigma^z})$ is simply obtained by replacing the spin variables $\mathbf{s}$ with the corresponding Pauli-Z matrices $\bm{\sigma^z}$. The coupling matrix defines a structure in the problem that can be represented by a weighted graph with $N$ vertices, connected by undirected edges $i\leftrightarrow{j}$ that have associated weights $J_{ij}, J_{ji}$. In this work, we study families of models in which the $J_{ij}$ coefficients are randomly drawn from a normal distribution $N(\mu=0,\sigma^2)$.    

This structure together with the magnetic field values defines families of optimization problems with different inner correlations and degeneracies. In the following subsections, we derive the interference amplitude for two scenarios that involve distinct energy level structure due to intrinsic global symmetries in the model. These differences necessitate a separate study of  both cases. Nevertheless, as explained below the slightly different behaviors of these models converge to a common expression of the QAOA probability amplitude distribution $|F(\mathbf{x})|^2$ in Eq.~\eqref{eq_multilevelinterfampl}. The two scenarios are represented by two well-known combinatorial optimization problems: Quadratic Unconstrained Binary Optimization (QUBO)~\citep{Kochenberger_2004,Kochenberger_2014}, with a non-degenerate energy spectrum, and the maximum cut (MaxCut)~\citep{Nannicini_2019,Sung_2020,Harrigan_2021}, which exhibits a global $\mathbb{Z}_2$ symmetry.

\subsubsection{Non-degenerate Ising Models}

The family of QUBO problems is composed of NP-Hard binary combinatorial optimization problems with the following associated cost function:
\begin{equation}
    E_{\textnormal{QUBO}}(\mathbf{x}) = 2 \sum_{i,j} x_i Q_{i j} x_j ,
\label{eq_QUBO}
\end{equation}
where $\mathbf{x} \in \{0, 1\}^N$, and $\mathbf{Q}$ is an $N$-by-$N$ square matrix. The mapping of Eq.~\eqref{eq_QUBO} to the Ising Hamiltonian in Eq.~\eqref{eq_isinghamiltonian}, $\mathbf{x} = \frac{1}{2}(1+\mathbf{s})$, leads to 
\begin{equation}
    \tilde{E}_{\textnormal{QUBO}} = 2 \sum_{i,j} x_i Q_{i j} x_j - \sum_{i,j} \frac{1}{2} Q_{i,j} = \frac{1}{2} \sum_{i,j=1}^N s_i  J_{ij} s_j + \sum_{i=1}^N h_i s_i ,
    \label{eq_QUBOtilde}
\end{equation}
where $\mathbf{J} = \mathbf{Q}$ and $h_{j}=\sum_i (Q_{ij}+Q_{ji})/2$. We consider $\mathbf{Q}$ matrices where the non-zero coefficients are randomly drawn from a standard normal distribution $N(\mu=0,\sigma^2=1)$. Note that the optimization of the function in Eq.~\eqref{eq_QUBOtilde} is then equivalent to that of Eq.~\eqref{eq_QUBO}. 

\begin{figure}[]
\begin{center}
\includegraphics[width=0.8\textwidth]{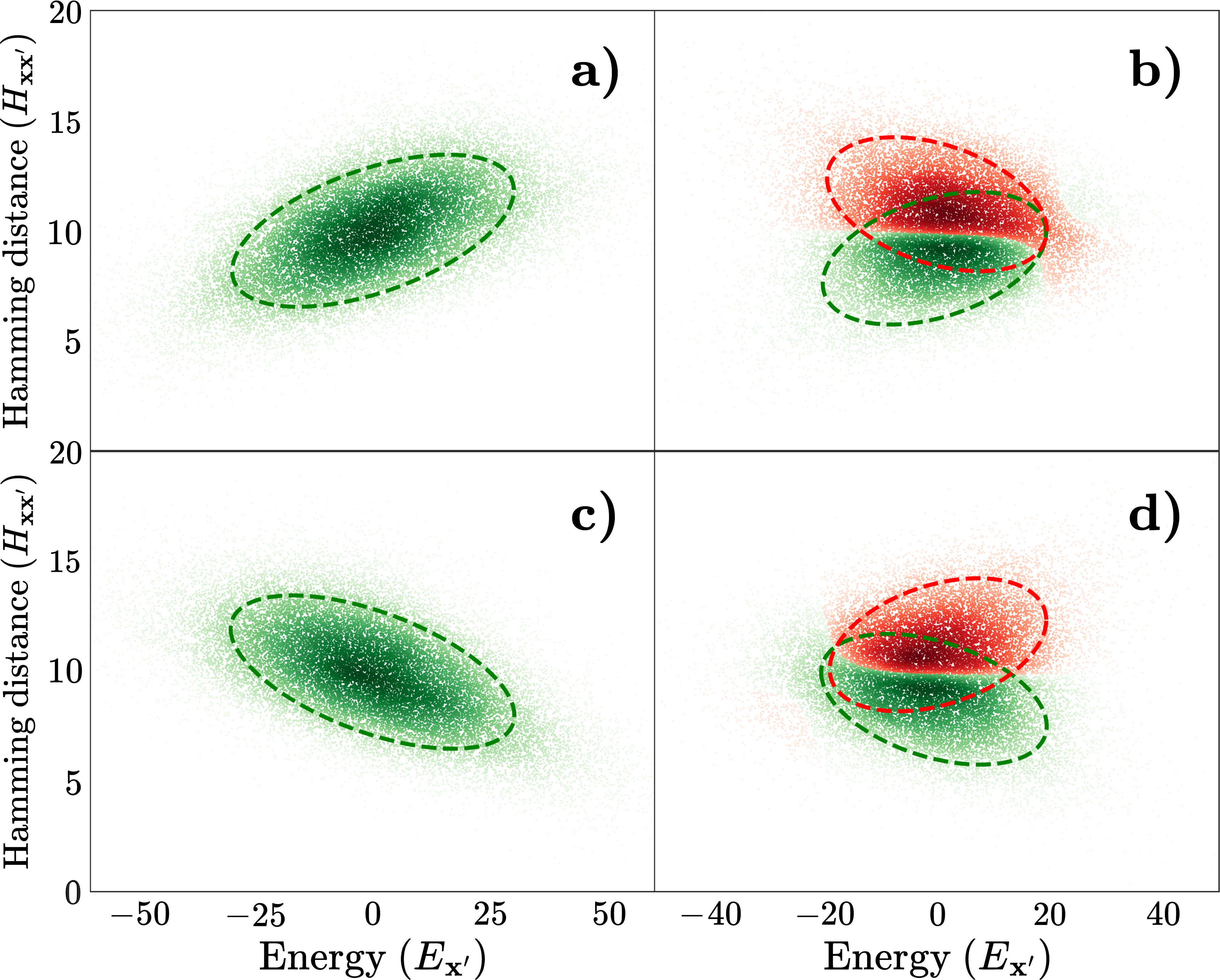}
\end{center}
\caption{Continuous approximation to the probability distribution $p(H,E; \mathbf{x})$ estimated by 25000 samples drawn from a kernel density estimation (KDE) on a single instance of a non-degenerate problem (a and c), and of a degenerate one (b and d). We plot the extreme cases when the reference state $\mathbf{x})$ is the ground state (a and b), and the highest energy state (c and d). We also fit the obtained distribution to a Gaussian mixture of one (a and c) or two (b and d) bivariate Gaussians, showing the clustering of the samples and the confidence ellipsoid over the distributions.}\label{Fig:HEdistribution}
\end{figure}

The cost function in Eq.~\eqref{eq_QUBOtilde} exhibits no global symmetries. In such non-degenerate situations, the eigenstates appear to be ordered from low to excited states, developing a unique probability distribution $p(H,E;\mathbf{x})$ that is centered at the center of the spectrum and shows a remarkable correlation between $H_\mathbf{xx'}$ and $E_\mathbf{x'}$. Such behavior can be well observed in Fig.~\ref{Fig:HEdistribution}. Here we plot the distribution in the Hamming distance-energy plane of 25000 samples drawn from a continuous approximation to $p(H,E;\mathbf{x})$ of a single instance of QUBO. In specific, we first estimate the continuous probability density function by a Kernel Density Estimation (KDE) on the actual discrete instance data (all $H_\mathbf{xx'}$ and $E_\mathbf{x'}$ pairs). Then we sample from such density function to obtain the plotted distributions. We also perform a fitting of the final distribution to a Gaussian mixture of one or two Gaussians using Variational Inference and plot the corresponding confidence ellipsoid. This fit was obtained with the \textit{BayesianGaussianMixture} method of the Python package \textit{scikit-learn}~\citep{scikit-learn}. We show the result for $\mathbf{x}$ being the ground state (a) and the highest energy state (c). For the sake of clarity, let us consider the upper-left plot, which illustrates the probability distribution $p(H, E; \mathbf{x})$ for the reference state $\mathbf{x}$ as the ground state. All alternative states $\mathbf{x'}$ possess higher energy values compared to $E_\mathbf{x}$. The difference $E_\mathbf{x'} - E_\mathbf{x}$ shows a potential correlation with the number of spin flips required to transition from $\mathbf{x}$ to $\mathbf{x'}$, i.e the Hamming distance $H_\mathbf{xx'}$. This positive correlation is evident through the presence of a Gaussian function oriented along a diagonal. On the contrary, opting for $\mathbf{x}$ as the highest energy state (bottom-left plot) reveals an opposite trend: the greater the number of spin flips, the lower the energy level. The Gaussian orientation depends on the covariance between $H$ and $E$,
\begin{equation}
    \sigma_{EH} = \dfrac{1}{2^{N}}\sum_{\mathbf{x'}} \left(H_{\mathbf{x}\mathbf{x'}} - \mathbb{E}[H] \right)\left(E_{\mathbf{x'}} - \mathbb{E}[E]\right),
    \label{eq_covEH}
\end{equation}
which is highly correlated with the energy of the reference state $E_{\mathbf{x}}$, as shown in Fig.4 of~\cite{Diez_Valle_2023}. In Eq.~\eqref{eq_covEH} $\mathbb{E}[\cdot]$ denotes the expected value of the variable. 

\jjgr{Our simulations confirm that, in the non-degenerate scenario, the probability distribution $p(H,E; \mathbf{x})$ resembles} a bivariate Gaussian with a correlation between variables defined by $\mathbf{x}$ and its rank in the energy spectrum:
\begin{equation}
  p(H,E; \mathbf{x})_{\rho}\approx
  \dfrac{1}{2\pi\sigma_E\sigma_H\sqrt{1-\rho^{2}}}
  \exp\left[-\dfrac{1}{2(1-\rho^{2})}
    \left(\left(\dfrac{E}{\sigma_E}\right)^2+\left(\dfrac{H-\mu_H}{\sigma_H}\right)^2-2\rho\dfrac{E(H-\mu_H)}{\sigma_E\sigma_H}\right)\right],
  \label{eq_bigaussian}
\end{equation}
where
\begin{equation}
\mu_E=\mathbb{E}[E]=0\,;\,\mu_H=\mathbb{E}[H]=\frac{N}{2}\,;\,\sigma_H= \frac{\sqrt{N}}{2}\,;\,\rho = \dfrac{\sigma_{EH}(\mathbf{x})}{\sigma_E\sigma_H}.
\label{eq_gaussianparameters}
\end{equation}
The spin model defines all the parameters of the bivariate Gaussian except the correlation parameter $\rho$, which encapsulates the whole dependence of the distribution on $\mathbf{x}$.

\begin{figure}
\begin{center}
\includegraphics[width=1\textwidth]{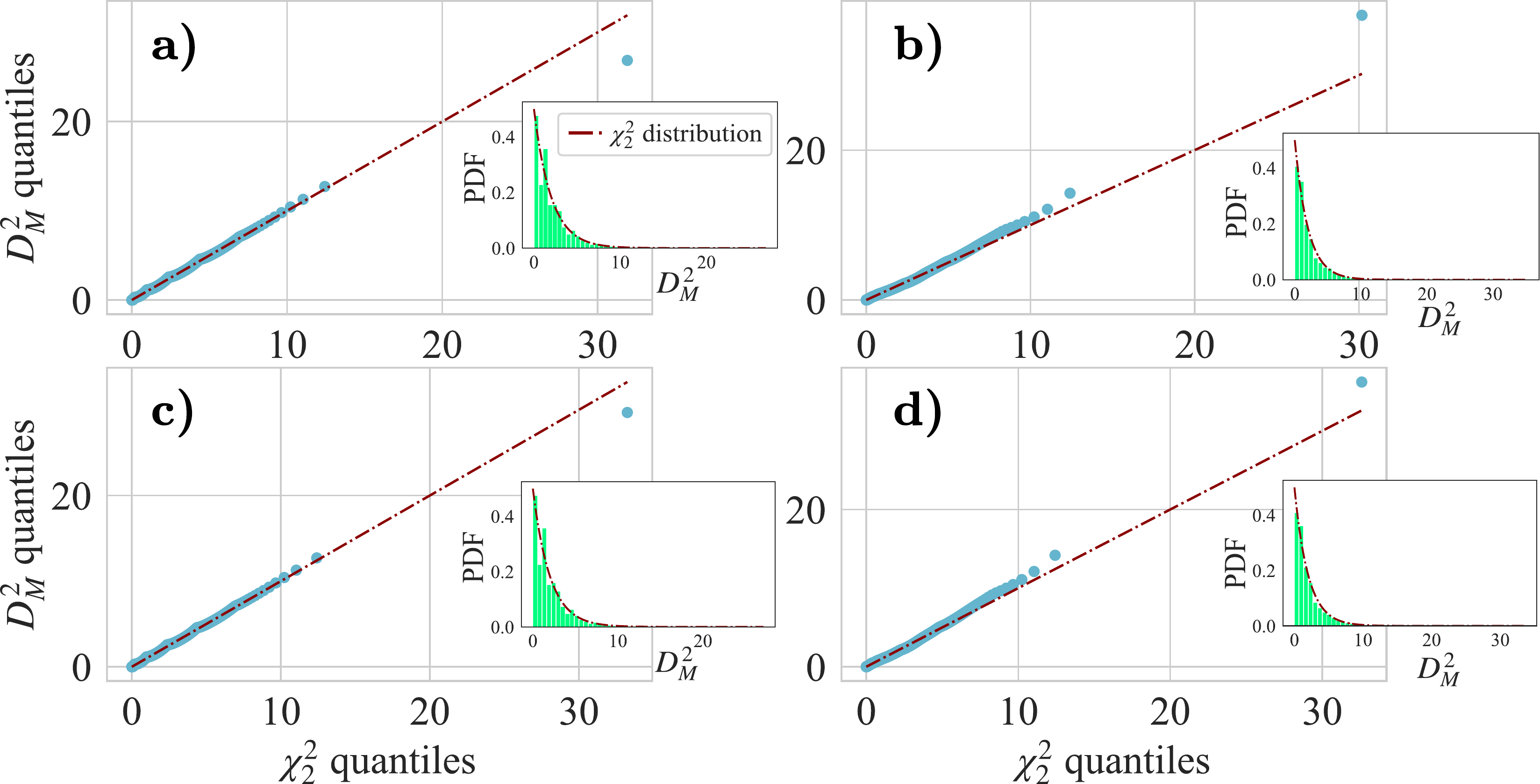}
\end{center}
\caption{Multivariate normality test to show that the structure of the energy levels of the Ising models studied resembles probability distributions $p(H,E;\mathbf{x})$ that can be defined by one or two continuous bivariate Gaussians. We plot the results for 1000 14-qubit random instances of QUBO (a,c) and MaxCut (b,d) when the reference state $\mathbf{x}$ is the ground state (a,b) and the highest energy state (c,d). For each instance of the problem, we calculated the Mahalanobis distances, $D_{M}$, of all ($H_{\mathbf{x}\mathbf{x'}}$,$E_{\mathbf{x'}}$) samples, and display the results of the obtained distributions. If such samples are compatible with a set of samples drawn from a bivariate Gaussian, the obtained Mahalanobis distances must follow the $\chi^2_2$ distribution. In the MaxCut case (b,d), a prior fit to a Gaussian Mixture was performed, so that the $\mathbf{x'}$ states were clustered into two Gaussians with corresponding means and covariance matrices. The smaller plots show the probability density function of $D_{M}^2$ as a histogram along with the theoretical $\chi^2_2$ distribution. The larger plots show quantile-quantile plots displaying 500 quantiles of $D_{M}^2$ and the same theoretical quantiles of the $\chi^2_{2}$ distribution. Over the scatter plot, we plot the straight line that would follow the points if $D_{M}^2$ and $\chi^2_2$ were described by the same distribution. We can see how about $99.8\%$ of the $D_{M}^2$ distribution is well described by the bivariate Gaussians, with the exception of $\sim0.2\%$ of outliers in the tail.}\label{Fig:MDdistribution}
\end{figure}

To test the hypothesis that $p(H,E; \mathbf{x})$ in Eq.~\eqref{eq_probHE} is well approximated by the continuous bivariate Gaussian distribution~\eqref{eq_bigaussian} we perform a graphical multivariate normality test on the ($H_{\mathbf{x}\mathbf{x'}}$,$E_{\mathbf{x'}}$) data of 1000 independent instances of QUBO. As in Fig.\ref{Fig:HEdistribution}, we analyze both the low and the high energy regimes with $\mathbf{x}$ the ground state and the highest energy state respectively. We use a technique based on the squared Mahalanobis distances between the sample points $\mathbf{s}=(H_{\mathbf{x}\mathbf{x'}},E_{\mathbf{x'}})$ and their averages $\bm{\mu}=(\mu_H,\mu_E)=(\mathbb{E}[H], \mathbb{E}[E])$ over all bit configurations $\mathbf{x'}$,    
\begin{equation}
    D_{M}(\mathbf{s}) = \sqrt{(\mathbf{s}-\bm{\mu})\Sigma^{-1}(\mathbf{s}-\bm{\mu})^T } = \dfrac{1}{\sqrt{1-\rho^2}}\left(\left(\dfrac{E}{\sigma_E}\right)^2+\left(\dfrac{H-\mu_H}{\sigma_H}\right)^2-2\rho\dfrac{E(H-\mu_H)}{\sigma_E\sigma_H}\right)^{\frac{1}{2}}
    \label{eq_mahalanobisdistance}
\end{equation}
where $\Sigma=\begin{pmatrix}
  \sigma_E & \sigma_{EH}  \\ \sigma_{EH} & \sigma_H
\end{pmatrix}$ is the covariance matrix, and $\rho = \frac{\sigma_{EH}}{\sigma_E\sigma_H}$. The Mahalanobis distance is a multivariate measure to quantify the distance between a point and a distribution~\citep{Mahalanobis1936}. Moreover, it is a useful tool to check when multidimensional data was sampled from a normal distribution, since the probability density $p$ of a set of normally distributed samples $\mathbf{s}$ in any dimension is entirely determined by the Mahalanobis distance,
\begin{equation}
   p(\mathbf{s}) = \frac{1}{\sqrt{(2\pi)^k|\Sigma|}} \exp\left[-\frac{1}{2}(\mathbf{s}-\bm{\mu})\Sigma^{-1}(\mathbf{s}-\bm{\mu})^T \right]  \Rightarrow p(D_{M}^2) =\frac{1}{\sqrt{(2\pi)^k|\Sigma|}}\exp\left[-\frac{D_{M}^2}{2}\right] . 
\end{equation}
Therefore, showing that the Mahalanobis distance $D_M(\mathbf{s})$, with $\mathbf{s}=(H_{\mathbf{x}\mathbf{x'}},E_{\mathbf{x'}})$, follows the chi-squared distribution with 2 degrees of freedom,
\begin{equation}
    \chi^2_2(x) = \frac{1}{2}\exp\left[-\frac{x}{2}\right] ,
\end{equation}
would demonstrate that $p(H,E)$ is compatible with a bivariate Gaussian distribution sampling (see Eq.~\eqref{eq_bigaussian}). Indeed, Fig.\ref{Fig:MDdistribution}a and Fig.\ref{Fig:MDdistribution}c shows the perfect agreement for at least $99.8\%$ of the spectrum.

Hence, returning to the probability amplitude of single-layer QAOA on non-degenerate Ising models, Eq.~\eqref{eq_amplint} together with the probability distribution in Eq.~\eqref{eq_bigaussian} leads to the following interference amplitude,
\begin{equation}
    |F(\mathbf{x})|^2 \propto \exp[Y],\mbox{ with} \,\,\, Y = -\gamma ^2 \sigma_E^2
    +\left(r^2-\frac{\pi^2}{4}\right)\sigma_H^2-2r\mu_H- \gamma  \pi \rho \sigma_E \sigma_H .
    \label{eq_anatampl}
\end{equation}

\subsubsection{Degenerate Ising Models}

The MaxCut problems are a family of combinatorial optimization problems consisting in minimizing the following objective function:
\begin{equation}
    E_{\textnormal{MaxCut}}(\mathbf{x}) = -2 \sum_{i,j} x_i Q_{i j} (1-x_j) ,
\label{eq_MAXCUT}
\end{equation}
with $\mathbf{x} \in \{0, 1\}^N$, and $\mathbf{Q}$ is an $N$-by-$N$ square matrix. Finding the minimum or an approximate solution very close to the minimum is known to be NP-Hard~\citep{Hastad_2001}. Again, we can map the binary cost function optimization in Eq.~\eqref{eq_MAXCUT} to an equivalent spin Hamiltonian optimization (Eq.~\eqref{eq_isinghamiltonian}),
\begin{equation}
    \tilde{E}_{\textnormal{MaxCut}} = -2 \sum_{i,j} x_i Q_{i j} (1-x_j) + \sum_{i,j} \frac{1}{2} Q_{i,j} = \frac{1}{2} \sum_{i,j=1}^N s_i  J_{ij} s_j ,
    \label{eq_MAXCUTtilde}
\end{equation}
where $\mathbf{J} = \mathbf{Q}$, and the magnetic field $\mathbf{h}$ cancels out. As in the non-degenerate case, the non-zero coefficients of $\mathbf{Q}$ are drawn from $N(\mu=0,\sigma^2=1)$. This class of problems includes the Sherrington-Kirkpatrick model~\citep{Sherrington_1975} when all vertices of the graph are connected, i.e. when the coupling matrix has no null coefficients.

In contrast to QUBO, the cost function in Eq.~\eqref{eq_MAXCUTtilde} exhibits a global $\mathbb{Z}_2$ symmetry that keeps invariant the energy under a global spin flip $\tilde{E}_{\textnormal{MaxCut}}(\mathbf{s}) = \tilde{E}_{\textnormal{MaxCut}}(-\mathbf{s})$, or what it is the same $\tilde{E}_{\textnormal{MaxCut}}(\mathbf{x}) = \tilde{E}_{\textnormal{MaxCut}}(1 -\mathbf{x})$. In the non-degenerate scenario (Eq.~\eqref{eq_QUBOtilde}) the presence of the magnetic field $\mathbf{h}$ breaks the symmetry. The existence of such global symmetries in the Hamiltonian results in the division of the Hilbert space into two or more distinct hierarchies of eigenstates. Note that if $\mathbf{x}^{a}=(x_1^{a},x_2^{a},\ldots )$ represents a ground state, then there is another ground state in the opposite sector ${\mathbf{x}}^{b}=1-{\mathbf{x}^{a}}$. Consequently, a single excited state $\mathbf{x'}$ can be seen now as arising from flipping spins in either of the ground states $\mathbf{x}^{a}$ or $\mathbf{x}^{b}$. Accordingly, an identical spin configuration exhibits significantly different Hamming distances to both ground states, while its energy remains unchanged. In specific, the Hamming distance reaches its maximum when considering the separation between the two degenerate ground states, where $H_{\mathbf{x}^{a}\mathbf{x}^{b}} = N$. Then, this symmetry implies that given any two states $\mathbf{x}$ and $\mathbf{x'}$, there is a unique alternative state $\mathbf{x''}$ such that $E_{x'}=E_{x''}$ and $H_{\mathbf{x}\mathbf{x'}} = N - H_{\mathbf{x}\mathbf{x''}}$.

This phenomenon results in the separation of the probability distribution into a sum of two \jjgr{distributions}, each measured with respect to one of the eigenstate hierarchies,
\begin{equation}
   p(H,E; \mathbf{x}) = p_{+}(H_{\mathbf{x},\mathbf{x'}},E_{\mathbf{x'}};\mathbf{x}) + p_{-}(H_{\mathbf{x},\mathbf{x'}},E_{\mathbf{x'}};\mathbf{x}),
   \label{eq_bigaussian_plusminus}
\end{equation}
where
\begin{align}
      p_{+}(H, E; \mathbf{x})
   & = \frac{1}{2^{N/2}}\sum_{\mathbf{x'}\in \mathcal{A}} \delta(H - H_\mathbf{xx'}) \delta(E - E_\mathbf{x'}),
   \label{eq_p_plus}\\
   p_{-}(H, E; \mathbf{x})
   & = \frac{1}{2^{N/2}}\sum_{\mathbf{x'}\in \mathcal{B}} \delta(H - H_\mathbf{xx'}) \delta(E - E_\mathbf{x'}),
   \label{eq_p_minus}
\end{align}
with $\mathcal{A}$, $\mathcal{B}$ two complementary subspaces in the bit configurations space that represent the two eigenstate hierarchies. Since each distribution $p_{\pm}(H_{\mathbf{x},\mathbf{x'}},E_{\mathbf{x'}};\mathbf{x})$ defines itself a non-degenerate Hilbert subspace, it is natural to expect that they individually behave similarly to the probability distribution of the non-degenerate scenario explained in the previous section. Indeed, as shown in Fig.\ref{Fig:HEdistribution}, these distributions resemble two shifted bivariate Gaussian distributions,
\begin{align}
  p_{+}(H,E;\mathbf{x})_{\rho_{+}}&\approx
  \dfrac{\exp\left[-\dfrac{1}{2(1-\rho_{+}^{2})}
    \left(\left(\dfrac{E}{\sigma_E}\right)^2+\left(\dfrac{H-\mu_{H}+h_0}{\sigma_H}\right)^2-2\rho_{+}\dfrac{E(H-\mu_{H}+h_0)}{\sigma_E\sigma_H}\right)\right]}{2\pi\sigma_E\sigma_H\sqrt{1-\rho_{+}^{2}}},
  \label{eq_bigaussian_plus}\\
  p_{-}(H,E;\mathbf{x})_{\rho_{-}} & = p_{+}(-H+2\mu_H,E;\mathbf{x})_{\rho_{+}}
   \\
  &\approx \dfrac{\exp\left[-\dfrac{1}{2(1-\rho_{+}^{2})}
    \left(\left(\dfrac{E}{\sigma_E}\right)^2+\left(\dfrac{H-\mu_{H}-h_0}{\sigma_H}\right)^2+2\rho_{+}\dfrac{E(H-\mu_{H}-h_0)}{\sigma_E\sigma_H}\right)\right]}{2\pi\sigma_E\sigma_H\sqrt{1-\rho_{+}^{2}}},
    \label{eq_bigaussian_minus}
\end{align}
where $\mu_E,\mu_H$ and $\sigma_{H}$ are the same as Eq.~\eqref{eq_gaussianparameters}, $h_0>0$ is a constant shift, and we have two separate correlation factors $\rho_{+}(\mathbf{x}) = - \rho_{-}(\mathbf{x}) \equiv \rho$. As in QUBO, we prove the Gaussian distribution of the eigenstates by a multivariate normality test based on the Mahalanobis distance. In this case, we first need to group the states in two clusters that represent $\mathcal{A}$ and $\mathcal{B}$ hierarchies. As previously explained, we do this by the fit of the sample points $\mathbf{s}=(H_{\mathbf{x}\mathbf{x'}},E_{\mathbf{x'}})$ to a mixture of two Gaussians so that we identify which states should belong to $p_{+}$ or $p_{-}$. Then we calculate the Mahalanobis distance of all samples using their corresponding mean and covariance estimated from the Gaussian mixture fit. Again, we find a good correspondence between the Mahalanobis distance and the $\chi_2^2$ distribution for the vast majority of the energy spectrum, concluding that $p_{\pm}(H,E;\mathbf{x})$ are well approximated by the Gaussian expresions in Eqs.~\eqref{eq_bigaussian_plus} and~\eqref{eq_bigaussian_minus}. 

As in the non-degenerate case, the correlation factors $\rho_{\pm}$ encapsulate the entire dependency in $\mathbf{x}$, but we now have two functions that influence one another. Eq.~\eqref{eq_interferenceamplitudes} along this $p_{+}$, $p_{-}$ interference leads to 
\begin{align}
    |F(\mathbf{x})|^2 &\propto \exp\left[Y'\right]\cdot\left(\cos{\left(h_{0}\pi+2r\gamma\rho\sigma_E\sigma_H\right)}+\cosh{\left(2h_{0}r-\gamma\pi\rho\sigma_E\sigma_H\right)}\right),\mbox{ with}
    \label{eq_anatampl_degenerate}\\
    Y' &\equiv -\gamma ^2 \sigma_E^2+\left(r^2-\frac{\pi^2}{4}\right)\sigma_H^2-2r\mu_H .
\end{align}
Therefore, we observe that the interference translates into a mixture of two exponentials together with an oscillatory term, 
\begin{equation}
     2\cosh{\left(2h_{0}r-\gamma\pi\rho\sigma_E\sigma_H\right)} = \exp{\left[\beta'\rho+2h_{0}r\right]} + \exp{\left[-\beta'\rho-2h_{0}r\right]},
\end{equation}
with $\beta'\equiv-\gamma\pi\sigma_E\sigma_H$. Nevertheless, except in the regime when $r = -\log(\tan \theta)\approx 0$ ($\theta\approx\pi/4$), one exponential is clearly dominant over the other due to the presence of the $h_0$-shift,
\begin{equation}
    \exp{\left[\beta'\rho+2h_{0}r\right]} + \exp{\left[-\beta'\rho-2h_{0}r\right]} \approx \exp{\left[\sgn\left(r\right)\beta'(\rho_{+}+\rho_{-})+\sgn\left(r\right)2h_{0}r\right]},
    \label{eq_expcosh_approx}
\end{equation}
where $\sgn(\cdot)$ is the sign function. Thus, the interference amplitude in Eq.~\eqref{eq_anatampl_degenerate} becomes
\begin{align}
    |F(\mathbf{x})|^2 &\propto \exp{[Y']}\cdot\exp{\left[\sgn\left(r\right)\beta'\rho\right]} = \exp[Y]
    \label{eq_anatampl_degenerate2},\mbox{ with}\\
    Y &= -\gamma ^2 \sigma_E^2
    +\left(r^2-\frac{\pi^2}{4}\right)\sigma_H^2-2r\mu_H-\sgn(r)\gamma\pi\sigma_E\sigma_H \rho 
   \label{eq_yaprox_deg}
\end{align}
Note that this formula is consistent with the non-degenerate scenario in Eq.~\eqref{eq_anatampl} except small corrections caused by the merging of eigenstate hierarchies (see Eq.~\eqref{eq_expcosh_approx}), and the role of $r$ that defines two different regimes for $\theta \in\left(0, \frac{\pi}{4}\right)$ and $\theta \in\left(\frac{\pi}{4},\frac{\pi}{2}\right)$.   

\section{QAOA thermal-like distributions}
\label{sec:QAOA thermal-like distributions}

\jjgr{The interferometric model, in combination with the approximate Gaussian correlations between energy and Hamming distance, is a powerful tool that allows us to approximate the probability distribution generated by the QAOA variational circuit. In the following discussion, we will indeed show that, with minimal assumptions, the single-layer QAOA state approximates a Boltzmann distribution with effective temperature determined by the $\gamma$ and $\theta$ angles, as shown in by~\citep{Diez_Valle_2023}.}

\jjgr{To achieve this goal, we analyze the single-layer QAOA interference amplitudes (Eqs.~\eqref{eq_anatampl} and~\eqref{eq_anatampl_degenerate2}) in energy space} 
\begin{equation}
    P(E) = \sum_{\mathbf{x}} \delta\left(E - E_{\mathbf{x}}\right)|F(\mathbf{x})|^2  . 
    \label{eq_probE}
\end{equation}
In order to examine this probability amplitude distribution of energy states, we should only pay attention to the terms of $|F(\mathbf{x})|^2$ that are influenced by the spin configuration $\mathbf{x}$ and its associated energy $E_{\mathbf{x}}$. In the previous section, we derived that the single-layer QAOA probability amplitude of Ising Hamiltonian eigenstates for both non-degenerate and degenerate spectra can be expressed as 
\begin{equation}
    |F(\mathbf{x})|^2 \propto \exp\left[-\gamma ^2 \sigma_E^2
    +\left(r^2-\frac{\pi^2}{4}\right)\sigma_H^2-2r\mu_H- \gamma  \pi \rho \sigma_E \sigma_H\right] ,
    \label{eq_unified_anatampl}
\end{equation}
where $\rho = \sigma_{EH}/(\sigma_E\sigma_H)$ with $\sigma_{EH} = \sgn(r)\sigma_{EH}$ in the degenerate scenario. As previously mentioned, the correlation factor captures the whole variability of Eq.~\eqref{eq_unified_anatampl} in energy space. All the other terms are set by the spin model and therefore are independent of the state $\mathbf{x}$. The only contributing part to the amplitude is 
\begin{equation}
-\gamma\pi \rho \sigma_E \sigma_H  = -\gamma \pi \sigma_{EH}(\mathbf{x}).
\end{equation}
Hence, in terms of the distribution in energy space, we can write the quantum probability amplitude as the following exponential,
\begin{equation}
|F(\mathbf{x})|^2 \propto e^{-\gamma \pi \sigma_{EH}(\mathbf{x})}.
\label{eq_amplcov}
\end{equation}

Let us recall that $\sigma_{EH}(\mathbf{x})$ represents the covariance between the Hamming distance from excited $\mathbf{x'}$ to the reference state $\mathbf{x}$ and their energy $E_{\mathbf{x'}}$ (Eq.~\eqref{eq_covEH}). The place of $\mathbf{x}$ in the energy spectrum is highly correlated with $\sigma_{EH}(\mathbf{x})$. In non-degenerate spaces, when $\mathbf{x}$ is the ground/highest-energy state it is more likely to find low/high energy states $\left((E_{\mathbf{x'}} - \mathbb{E}[E])<0 / (E_{\mathbf{x'}} - \mathbb{E}[E])>0\right)$ close to the reference state $\left((H_{\mathbf{x}\mathbf{x'}} - \mathbb{E}[H])<0\right)$ and high/low energy states far $\left((H_{\mathbf{x}\mathbf{x'}} - \mathbb{E}[H])>0\right)$ leading to a high positive/negative covariance. In the degenerate scenario, we have the same behavior for subspace $p_{+}$ and the exact opposite for the complementary subspace $p_{-}$ $\left(H_{\mathbf{x}\mathbf{x'}}\rightarrow -H_{\mathbf{x}\mathbf{x'}}+2\mu_H\right)$. This intuition translates into an evident  $\sigma_{EH}-E_{\mathbf{x}}$ correlation that can be numerically observed in Fig. 4 of~\citep{Diez_Valle_2023}. This dependence can be expressed as the sum of a linear function and a stochastic value $\omega$ with zero mean
\begin{equation}
    \sigma_{EH}(\mathbf{x}) = -c\cdot E_{\mathbf{x}} \pm\omega,
    \label{eq_covEcorr}
\end{equation}
where $c\in\mathbb{R}>0$. In spite of the presence of the random term $\omega$, the trend in Eq.~\eqref{eq_covEcorr} is noticeable. Therefore, introducing Eq.~\eqref{eq_covEcorr} in Eq.~\eqref{eq_amplcov} leads to a thermal-like probability amplitude distribution for the Ising models,
\begin{equation}
|F(\mathbf{x})|^2 \propto e^{-\beta E_\mathbf{x}\pm  \beta \omega/c},
\label{eq_amplasbolt}
\end{equation}
where $\beta = c\pi\gamma$ for non-degenerate problems, and $\beta = \sgn(r)c\pi\gamma$ for degenerate ones, plays the role of the inverse temperature. Then the probability distribution in energy space can be expressed as,
\begin{equation}
    P(E) \sim d(E)e^{-\beta E},
\end{equation}
where $d(E)$ is the density of states that in Ising models resembles a normal distribution centered in intermediate energies.

\begin{figure}[]
\begin{center}
\includegraphics[width=1\textwidth]{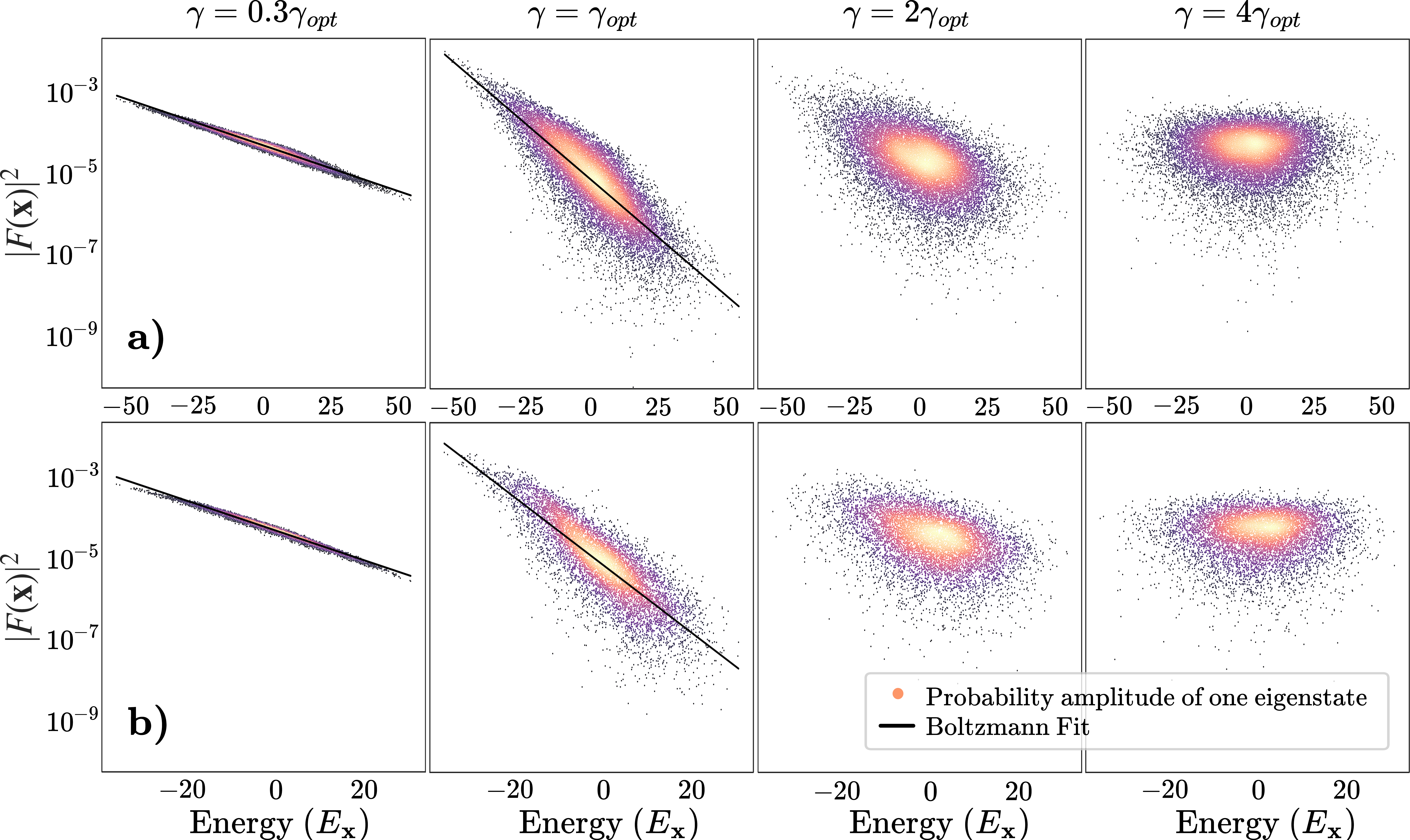}
\end{center}
\caption{Evolution of the single-layer QAOA probability amplitude distribution as we modify the angle $\gamma$ with optimal $\theta$. We plot the probabilities $|F(\mathbf{x})|^2$ versus their energies $E_{\mathbf{x}}$ for a single random instance of a 14-qubits (a) QUBO ($\gamma_{opt}=-0.14$, $\theta_{opt}\approx\pi/6$), and (b) MaxCut ($\gamma_{opt}=-0.10$, $\theta_{opt}\approx\pi/8$) Hamiltonian. The optimal angles ($\gamma_{opt},\theta_{opt}$) are those that minimize the mean energy $\langle E\rangle = \sum_{\mathbf{x}}|F(\mathbf{x})|^2 E_{\mathbf{x}} $. Note how a Boltzmann distribution with perturbations is apparent for $\gamma \leq \gamma_{opt}$.}\label{Fig:Boltzmannevolution_gamma}
\end{figure}

\begin{figure}[]
\begin{center}
\includegraphics[width=1\textwidth]{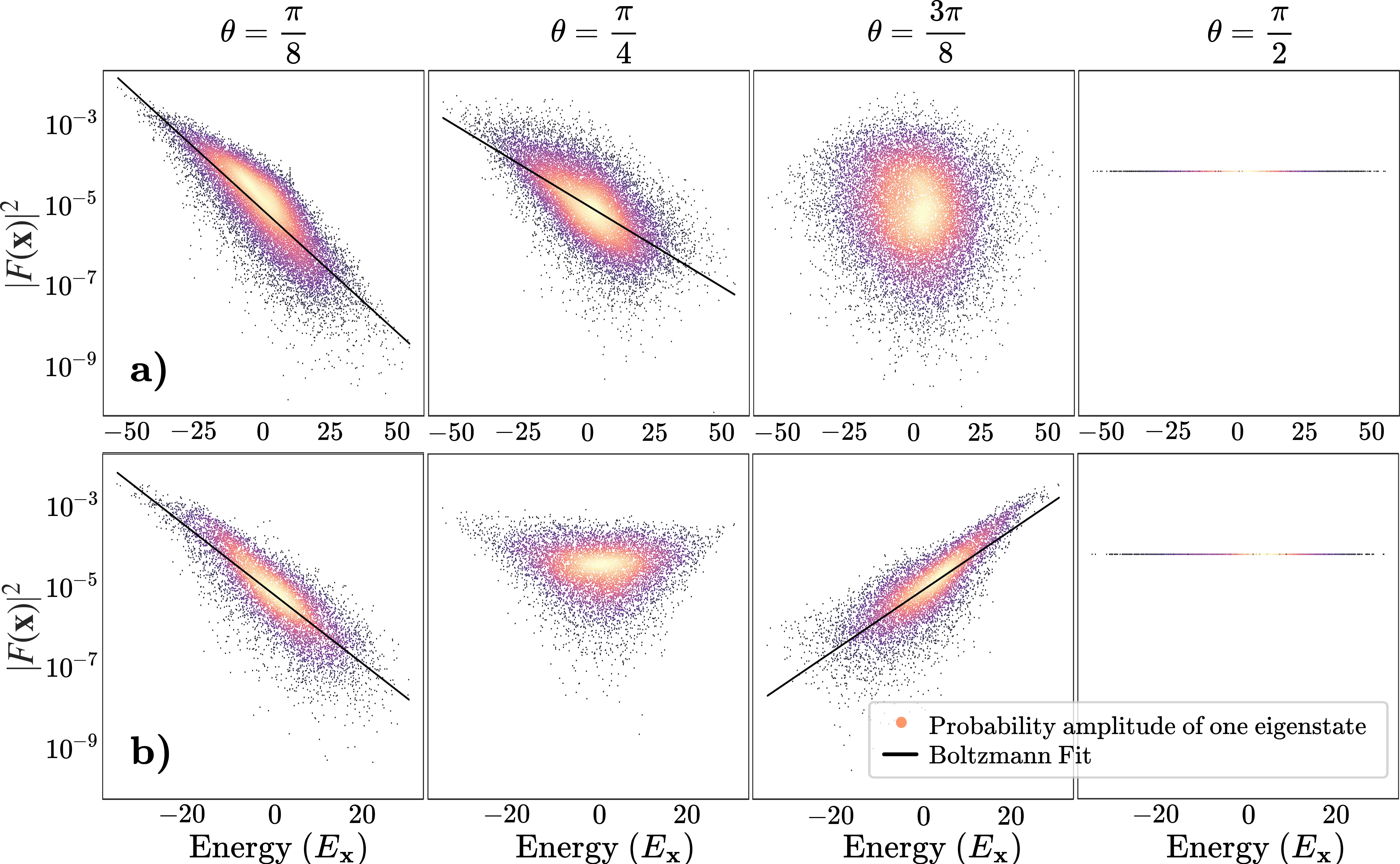}
\end{center}
\caption{Single-layer QAOA probability amplitude distribution with optimal $\gamma$ in different regimes of the angle $\theta$. We plot the probabilities $|F(\mathbf{x})|^2$ versus their energies $E_{\mathbf{x}}$ for a single random instance of a 14-qubits (a) QUBO ($\gamma_{opt}=-0.14$, $\theta_{opt}\approx\pi/6$), and (b) MaxCut ($\gamma_{opt}=-0.10$, $\theta_{opt}\approx\pi/8$) Hamiltonian. The optimal angles ($\gamma_{opt},\theta_{opt}$) are those that minimize the mean energy $\langle E\rangle = \sum_{\mathbf{x}}|F(\mathbf{x})|^2 E_{\mathbf{x}} $.}\label{Fig:Boltzmannevolution_theta}
\end{figure}

The Boltzmann distribution in Eq.~\eqref{eq_amplasbolt} together with the random fluctuations $\omega$ manifests in regimes of optimized parameters, but the distribution may be modified by manipulating the angles $\theta\in (0,\pi/2)$ and $\gamma$ (see Figs.\ref{Fig:Boltzmannevolution_gamma} and \ref{Fig:Boltzmannevolution_theta}). By the analysis of Eq.~\eqref{eq_amplasbolt} and  previous expressions, additional conclusions can be drawn about the shape of the distribution. 

\begin{itemize}
    \item The angle $\gamma$ controls the direction of the optimization. For specific $\gamma$ regimes, the single-layer QAOA probability amplitude distribution on Ising models resembles a thermal distribution at temperature $T=\beta^{-1}=\frac{1}{c\pi|\gamma|}$ such that the ground state of the system aligns with the peak amplitude. Switching the sign of this angle is the same as changing the sign of the energy, and therefore of the temperature $T=\frac{-1}{c\pi|\gamma|}$. These negative temperature states increase the probability of the highest excited state. 

    \item The degenerate Ising models exhibit antisymmetric behavior at the angle $\theta$ since $T=\frac{\sgn(r)}{c\pi|\gamma|}$, with $r=-\log(\tan\theta)$. Therefore, for $\theta < \pi/4$ and $\theta > \pi/4$ the pure QAOA state resembles a thermal-like distribution with positive/negative temperature. Eq.~\eqref{eq_anatampl_degenerate} shows that when $\theta\approx\pi/4$ the thermal type  distribution disappears and we find a mixture of two Boltzmann exponentials with opposite temperatures. This behavior can be well observed in Fig. \ref{Fig:Boltzmannevolution_theta}b. For non-degenerate Hamiltonians, in the $r \ll 0$ regime the $r^2\sigma_H^2$ and $-2r\mu_H$ terms of Eq.~\eqref{eq_unified_anatampl} increase the noise and the Boltzmann distribution vanishes.     

    \item The Boltzmann distribution is apparent for a finite interval of angles $\gamma \in (0,\gamma_c]$. The lowest temperature $T$ is reached near $\gamma \approx \gamma_{opt}$, where $\gamma_{opt}$ is the angle that minimizes the mean energy in the variational principle. When $|\gamma|>\gamma_c$, the Boltzmann term $\gamma\pi\rho\sigma_E\sigma_H$ becomes marginal in comparison with $-\gamma^2\sigma_E^2$ in Eq.~\eqref{eq_unified_anatampl}, so that $|F(\mathbf{x})|^2\approx cte$.

    \item For $\gamma < \gamma_c$, we observe that $\beta$ grows linearly with the angle $\gamma$, consistent with the numerical results of~\citep{Diez_Valle_2023}. 

\end{itemize}

\section{Outlook}
\label{sec:Outlook}

Sampling from complex probability distributions is a valuable computational task, both for its difficulty and its broad applicability. Quantum states projected on the computational basis are in essence classical probability distributions and measuring them is the same as sampling from such distributions. Therefore, a quantum computer can be roughly seen as a machine capable of creating exotic or suitable probability distributions thanks to quantum phenomena. Indeed, the first claim of \textit{quantum supremacy}~\citep{Preskill_2018} was performed on a 
sampling problem~\citep{Arute_2019}, and this task has been also envisioned as one of the candidates to show a practical quantum advantage in the near term~\citep{Wu_2021,Zhong_2021,Layden_2023}.  

In this context, the Boltzmann distributions of Ising models could be good candidates to show such quantum advantage for two reasons. First, these distribtuions are classically intractable at low-temperature regimes. The most popular strategy to simulate thermal distributions is the use of Markov Chain Monte Carlo algorithms (MCMC) which guarantee convergence to the Boltzmann distribution at a given temperature $T$. While these methods work very well for relatively high temperatures, the number of iterations needed to converge scales exponentially when the temperature is too low. Much effort has been done to determine the range of temperatures that ensure rapid mixing and therefore polynomial convergence of MCMC. To the best of our knowledge, the state-of-the-art theoretical bound indicates that MCMC always converges in polynomial time to the thermal distribution of an Ising model at a temperature higher than $\Vert{\mathbf{J}}\Vert$~\citep{eldan_2021}, although practical realizations and state-of-the-art methods (e.g. parallel tempering, population annealing ...) might overcome this threshold.~\citep{Diez_Valle_2023} showed that the single-layer QAOA already approximates Boltzmann distributions at temperatures beyond this theoretical MCMC bound.

Second, there is a wide range of fields that would be impacted by an improvement in Boltzmann distribution sampling. In statistical mechanics, 
sampling from this distribution is crucial for simulating physical systems at thermal equilibrium and for computing observables such as magnetization in Ising models. Furthermore, machine learning makes use of this distribution in unsupervised learning techniques known as Boltzmann Machines~\citep{ackley_1985}. Combinatorial optimization is another field of interest since some algorithms employ Boltzmann distribution sampling at decreasing temperatures as a subroutine to find the minimum of a cost function~\citep{SimulatedAnnealing1983}.

The thermal-like distributions of the single-layer Quantum Approximate Optimization algorithm reveal a nice connection between quantum algorithmics and statistical physics that can help to gain a better understanding of the behavior of these ansätze. But in a broader perspective, the QAOA sampling from approximate Boltzmann distributions, also known as Gibbs sampling, may be extended to the multi-layer scenario with an improvement in the achievable temperature~\citep{Lotshaw_2022}, or to QAOA mixed-state ansätze to train unsupervised learning models as implemented in~\citep{Verdon_2019}. Furthermore, since QAOA resembles a trotterized approximation to an adiabatic quantum evolution, this picture might be expanded to the infinite depth scenario $p\rightarrow\infty$ and the engineering of general time-dependent adiabatic passages.

Additionally, the approximate Boltzmann distributions of single-layer QAOA states present collateral implications in quantum information theory. For example, this behavior makes them suitable as warm initial states for more complex ansätze as has been demonstrated in~\citep{leontica2023exploring}. Another recent work~\citep{Sud_2022} shows how a tight dependence between the energy distribution of the spin model and the final probability amplitude of the QAOA states allows a more efficient classical heuristic optimization of the QAOA parameters. The Boltzmann distribution unambiguously connects the energy of the eigenstates with their amplitudes, thus providing further arguments and explanations on these heuristics.

\jjgr{We are confident that this study, in combination with earlier work by~\citep{Diez_Valle_2023}, are just two initial works in a new field in which variational and other types of circuits are analyzed from a physical perspective, understanding not only their computational power but potentially significant physical insight behind the quantum computer's dynamics.}

\section*{Acknowledgments}
This work was supported by  European Commission FET Open project AVaQus Grant Agreement 899561, the Proyecto Sinérgico CAM 2020 Y2020/TCS-6545 (NanoQuCoCM), the Spanish CDTI through Misiones Ciencia e Innovación Program (CUCO) under Grant MIG-20211005, and CSIC Interdisciplinary Thematic Platform (PTI) Quantum Technologies (PTI-QTEP).

\bibliographystyle{Frontiers-Harvard} 
\bibliography{bibliography}


\end{document}